\documentclass[a4paper,11pt]{article}
\usepackage{jinstpub} 
\usepackage{lineno}
\usepackage{float}

\title{\boldmath Photon Classification with Gradient Boosted Trees at CLAS12}

\author[a,1]{G. Matousek\note{Corresponding author.}}
\author[a,b]{, A. Vossen}
\affiliation[a]{Department of Physics, Duke University,\\
120 Science Drive, Durham, NC 27708, USA}
\affiliation[b]{Thomas Jefferson National Accelerator Facility,\\
12000 Jefferson Ave., Newport News, VA 23606, USA}

\emailAdd{gregory.matousek@duke.edu}

\abstract{Dihadron semi-inclusive deep inelastic scattering (SIDIS) of 10.6 GeV longitudinally polarized electrons off the proton has been measured using the CLAS12 detector at Jefferson Lab. Two separate channels, $\pi^+\pi^0$ and $\pi^-\pi^0$, were analyzed, requiring the reconstruction of diphoton pairs. In this analysis, we addressed the problem of false neutral particles being reconstructed by CLAS12's event builder, polluting the otherwise physical combinatorial background underneath the $\pi^0$ peak. A photon classifier using a Gradient Boosted Trees (GBTs) architecture was trained with Monte Carlo simulations to reduce the amount of background $\pi^0$'s. We show that the nearest-neighbor features learned by the model lead to a substantial increase in signal vs. background discrimination compared to previous CLAS12 $\pi^0$ analyses. The machine learning approach recovers several times more dihadron statistics for the dataset.}

\begin{document}
\maketitle
\flushbottom

\section{Introduction}
\indent Our understanding of the spin structure of nucleons has seen several advancements with the help of recent deep inelastic scattering (DIS) experiments. In semi-inclusive DIS (SIDIS), a single hadron is identified in the final state along with the reconstructed scattered electron. Measured asymmetries in the azimuthal angle, $\phi_h$, of these hadrons, with respect to either or some combination of the beam and target spin reveal the presence of spin-dependent parton distribution functions (PDFs) within nucleons. Extracting these PDFs in single hadron SIDIS involves modeling the intrinsic transverse momentum $k_T$ dependence of the partons. By studying the azimuthal modulations of dihadron SIDIS, these model-dependent approaches towards accessing PDFs are avoided. This is because the two hadron state has an additional degree of azimuthal freedom, $\phi_R$, which allows certain PDFs to be accessed without convolutions over transverse momentum space as $k_T$ is integrated out. In particular, the elusive twist-3 collinear PDF $e(x)$ can be extracted in such a way by measuring the $(\phi_h,\phi_R)$ asymmetries in dihadron SIDIS.\\
\indent In addition to offering a more targeted approach for measuring spin-sensitive PDFs, dihadron SIDIS allows us to study more complex spin-related effects in hadronization. At leading twist, dihadron fragmentation functions (DiFFs) encode the probability for a quark to fragment into a hadron pair $h_1h_2$ \cite{Gliske_Bacchetta_Radici_2014}. The DiFF $G_{1}^{\perp}$, which describes the quark helicity dependence on dihadron production, has had little experimental investigation up to this point. \\
\indent Previous beam spin asymmetries of $ep\rightarrow e\pi^+\pi^-X$ SIDIS dihadrons were measured using the CLAS12 detector system \cite{Hayward_Dilks_Vossen}. For our study, we extend this analysis using the same dataset to $\pi^+\pi^0$ and $\pi^-\pi^0$ dihadrons. Exploring the same asymmetries with different final states allow us to explore the isospin dependence of our fragmentation functions (see \cite{Matevosyan_Kotzinian_Thomas_2017L,Matevosyan_Kotzinian_Thomas_2018}), as well as PDFs such as $e(x)$ with different systematics.
\section{Experiment}
Dihadron $\pi^\pm\pi^0$ SIDIS was measured at Jefferson National Lab using the CLAS12 (CEBAF Large Acceptance Spectrometer 12-GeV) detector system \cite{Burkert_Elouadrhiri_Adhikari_Adhikari_Amaryan_Anderson_Angelini_Antonioli_Atac_Aune_etal._2020}. The data set corresponds to the Run Group A (RG-A) configuration which operated between Spring 2018 and Spring 2019. The experiment scattered a longitudinally polarized electron beam in the energy range of $10.2-10.6$ GeV with a run-by-run polarization of $86-89$\% on a fixed, unpolarized liquid hydrogen target. \\
\indent At CLAS12, a sector-based electron trigger is used to select inclusive scattering events, followed by a skimming procedure to extract candidate DIS events (ex: $Q^2>1$ GeV$^2$). Forward detectors ($5<\theta<35^\circ$) including drift chambers and electromagnetic calorimeters provide identification and four-momentum reconstruction of our electrons, charged pions, and photons. 

\section{Machine Learning}
\begin{figure}
\centering
\includegraphics[width=0.35\textwidth]{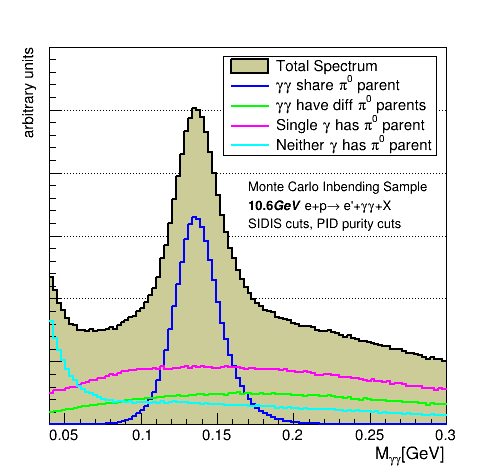}
\caption{Origin of diphoton signal and background distributions from reconstructed CLAS12 Monte Carlo. SIDIS cuts correspond to $Q^2>1$ GeV$^2$, $y<0.8$, and $W>2$ GeV. }
\label{fig:diphoton1}
\end{figure}

\begin{figure}
\centering
\includegraphics[width=0.35\textwidth]{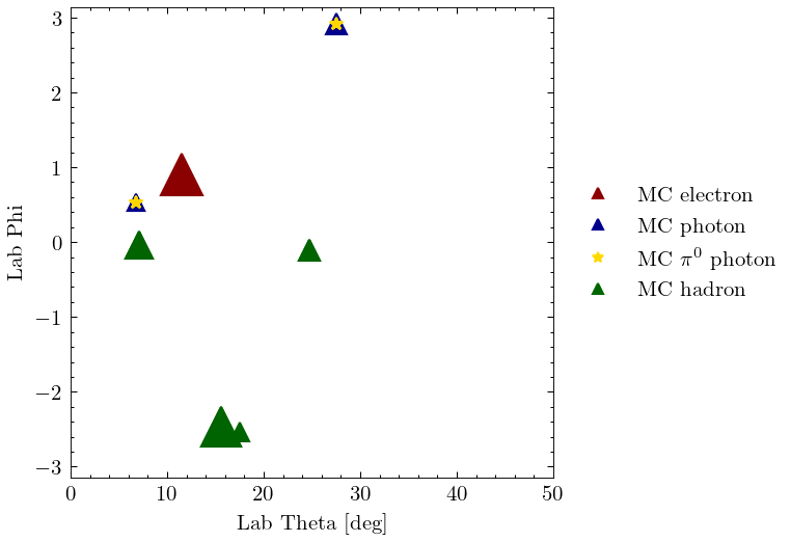}
\qquad
\includegraphics[width=0.35\textwidth]{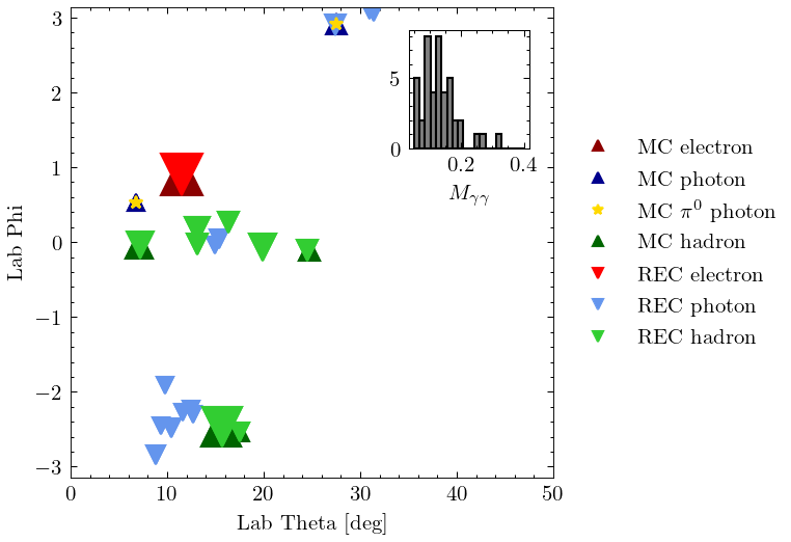} 
\caption{A sample Monte Carlo event drawn in ($\theta$,$\phi$) space. $\theta$ is the scattering angle of the particle off the beamline, and $\phi$ is the azimuthal angle. Marker size roughly scales with particle energy. On the left shows only the Monte Carlo particles, whereas the right overlays particle reconstructed by the Event Builder}
\label{fig:dihadronEvents}
\end{figure}

\indent The main challenge encountered during the $\pi^\pm\pi^0$ analysis was the abundance of non-physical $\gamma\gamma$ background beneath the $\pi^0$ peak. This background can be seen in Fig. \ref{fig:diphoton1}, where a large fraction of photons beneath the peak do not have $\pi^0$ parent (i.e. are not combinatorial backgrounds). At CLAS12, the absence of an ionized track in the drift chambers pointing towards a non-negligible energy deposition in the ECal is indicative of a neutral particle candidate ($\gamma$ or $n$, which are separated using timing). The particle reconstruction algorithm implemented across many CLAS12 analyses, called the Event Builder, incorrectly assigns background calorimeter clusters to physical photons from the event. The calorimeter clustering also tends to mistakenly "split" one local hotspot into several particles, as opposed to merging them into one single particle. The nature of these false neutrals is shown in a sample reconstructed Monte Carlo SIDIS event in Fig. \ref{fig:dihadronEvents}. Despite only 2 true event photons, 11 are reconstructed, with a majority of them grouped near other neutrals, or near other energetic hadrons.

\begin{figure}
\centering
\includegraphics[width=0.35\textwidth]{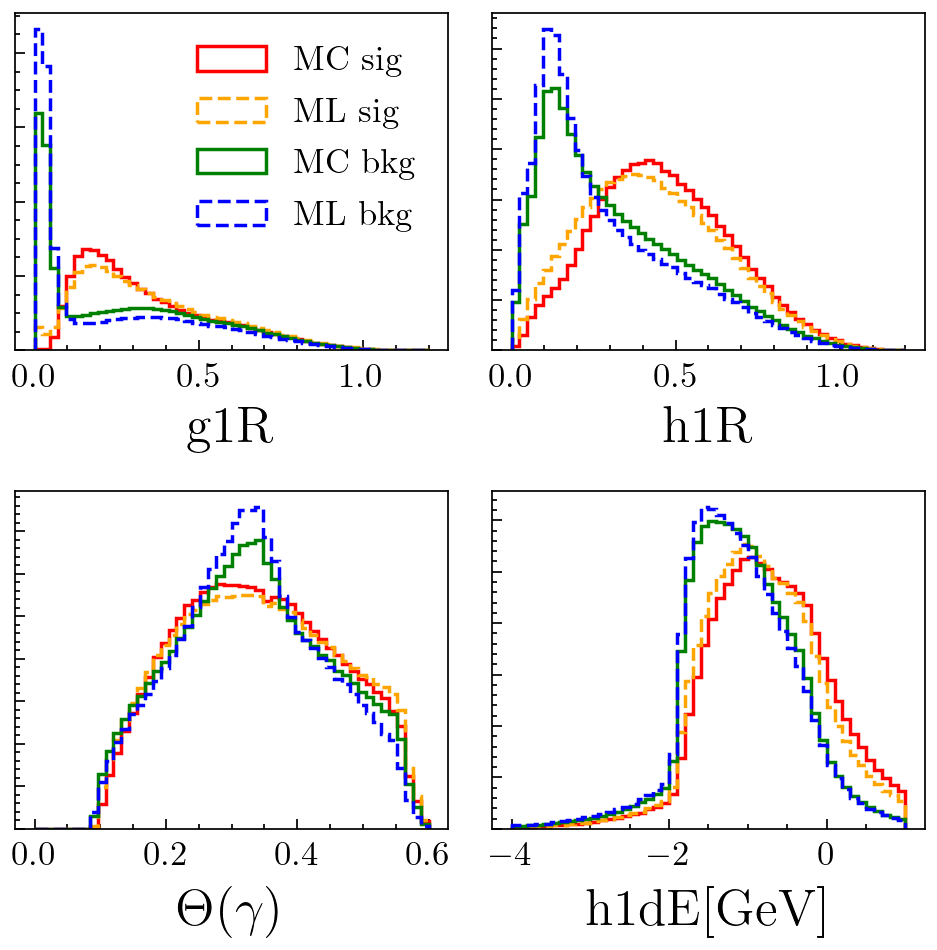}
\qquad
\includegraphics[width=0.35\textwidth]{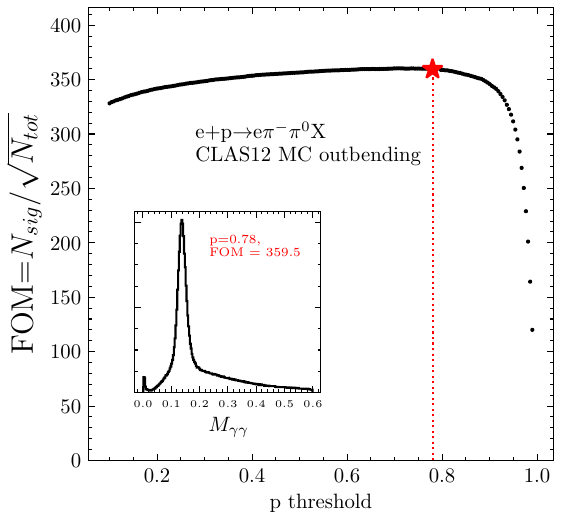}
\caption{\textbf{Left:} Feature distributions between true Monte Carlo (solid) and ML predictions. The four shown are the nearest photon neighbor relative angle (top left), nearest hadron neighbor relative angle (top right), photon scattering angle (bottom left) and nearest hadron neighbor relative energy (top right). \textbf{Right:} Figure of merit extraction for optimal $p$-threshold cut of the GBT model.}
\label{fig:4grid}
\end{figure}

\indent In this work, a machine learning model was developed to solve the non-physical diphoton background problem. The architecture we decided upon was Gradient Boosted Decision Trees (GBTs) classifier which was implemented using the open-source python package \texttt{CatBoost} \cite{catboost}. For each reconstructed photon in an event, the model was trained to classify if the photon was physical (i.e. originated from the SIDIS event). The training sample was collected by taking photons from reconstructed Monte Carlo and flagging those that had a generated Monte Carlo counterpart as physical (signal). The 16 input features of the model are split into 3 categories: event-wide (such as the number of photons in the event), those intrinsic to the photon (such as its $E_{\mathrm{tot}}$, $\theta$, and $E_{\mathrm{PCAL}}$), and, the most critical to the model's success, those intrinsic to its nearest neighbors. In particular, it was predicted and later confirmed through testing that the angular distance between the photon and its $k$-nearest charged hadrons, neutrals, and event electron were essential for accurate discrimination of signal vs. background. The discriminating power of several of these features are shown in Fig. \ref{fig:4grid}, shown alongside a figure of merit scan for determining the optimal $p$-threshold for the GBT model. Since we do not use the relative energy between neighboring photons, we prevent the model from learning the resonant structure of the $\pi^0$ peak. A final model was developed using 1,000 trees with a maximum depth of 10, adhering to a symmetric growth policy. A learning rate of 0.1 was applied to moderate gradient descent, and log loss was evaluated on a validation set split 75/25 with the training data. Log loss was recorded for each of the 1,000 generations, and the model with the lowest validation loss was selected to minimize overfitting.
\section{Results}
\begin{figure}
  \centering
  \includegraphics[width=0.325\textwidth]{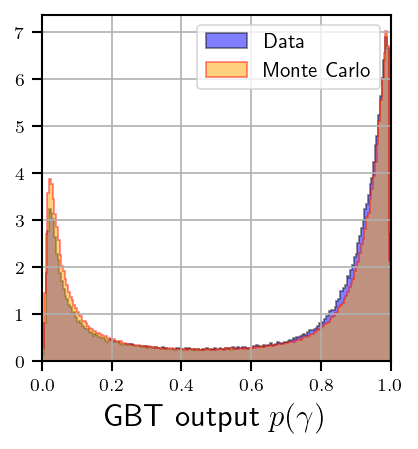}
  \qquad
  \includegraphics[width=0.325\textwidth]{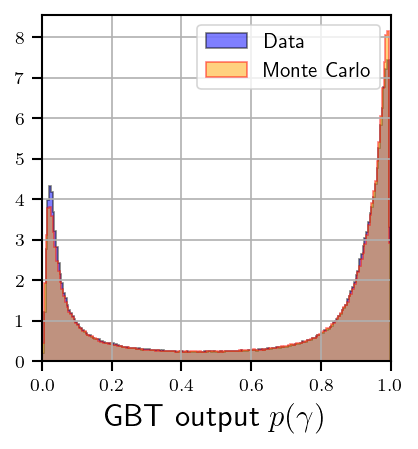}
  \caption{Output histograms of the trained GBTs when applying them on data and Monte Carlo. \textbf{Left:} $\pi^+\pi^0$ dihadrons. \textbf{Right:} $\pi^-\pi^0$ dihadrons.}
  \label{fig:pofx}
\end{figure}

\begin{figure}
  \centering
  \includegraphics[width=0.35\textwidth]{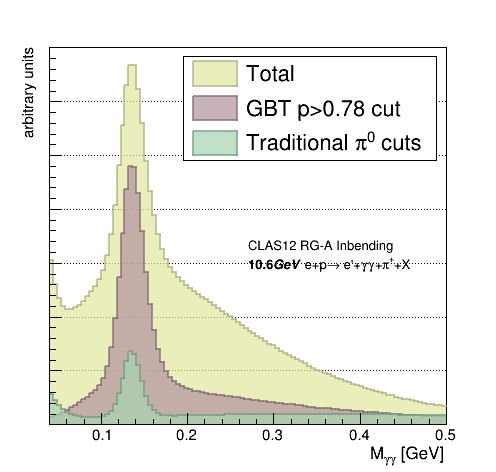}
  \qquad
  \includegraphics[width=0.35\textwidth]{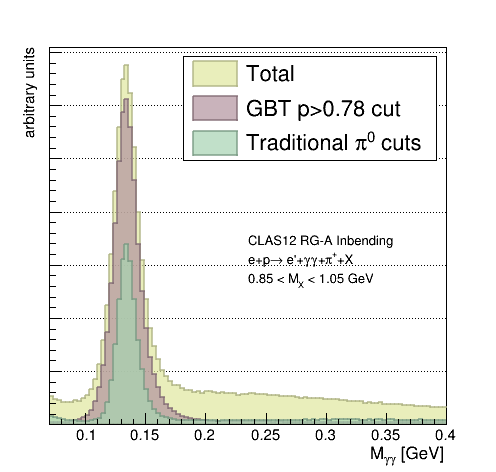}
  \caption{Diphoton spectra of SIDIS $\pi^+\pi^0$ dihadrons. \textbf{Left:} No exclusivity cut applied. \textbf{Right:} Exclusivity cut applied.}
  \label{fig:finalMgg}
\end{figure}

\indent In Fig. \ref{fig:pofx} we overlay the GBT model output when applied to data and Monte Carlo photons. The agreement of these distributions indicates the similarity of the data and Monte Carlo input feature space, supporting the claim that the model can be effectively used to classify signal data photons. In Fig. \ref{fig:finalMgg} we show the diphoton spectra of SIDIS $\pi^+\pi^0$ dihadrons. For each subfigure, we plot the distribution before applying any photon purity cuts, after using CLAS12 standard photon purity cuts (in this case, $E_{\gamma}>0.6$ GeV), and after using our machine learning approach. On the right figure, we apply an additional exclusivity cut on the final state missing mass to select exclusive $ep\rightarrow e\rho^+(n)$ events. We see that our machine learning model reduces the background fraction of $\gamma\gamma$ pairs underneath the $\pi^0$ peak while maintaining a significantly large amount signal $\pi^0$'s. The figure of merit for the GBT approach is 360, compared to the much smaller 182 figure of merit for the traditional cuts. In the exclusive case, where we expect no combinatorial diphoton background, the machine learning model's background diphoton distribution is flat near zero. Overall, we've shown that a GBT model trained on nearest-neighbor features tackles the false neutral backgrounds at CLAS12, providing a more statistically significant data sample for our dihadron study. 
\begin{acknowledgments}
This research was supported by the U.S. Department of Energy (DOE) and the National Science Foundation (NSF). We extend our gratitude to the DOE and NSF for their financial support, which made this work possible.
\end{acknowledgments}
\appendix


\bibliographystyle{JHEP}
\bibliography{main.bib}






\end{document}